\documentclass[twocolumn,showpacs,pra]{revtex4}
\usepackage{graphicx}

\newcommand{\ket}[1]{|{#1}\rangle}

\begin{document}
\title{Polarization Entangled {\it W} State using Parametric Down-Conversion}

\author{Takashi Yamamoto}
\email{yamamoto@koryuw01.soken.ac.jp}
\author{Kiyoshi Tamaki}
\author{Masato Koashi}
\author{Nobuyuki Imoto}
\homepage{http://www.soken.ac.jp/quantum/index.html}
\address{CREST Research Team for Interacting Carrier Electronics, School of Advanced Sciences,\\ The Graduate University for 
Advanced Studies (SOKENDAI), Hayama, Kanagawa, 240-0193, Japan
}

\date{\today}

\begin{abstract}
An experimental scheme for preparing a polarization entangled {\it W} states from four photons emitted by parametric down-conversion is proposed. We consider two different configurations and a method of improving the yield by using single photon sources. In the proposed scheme, one uses only linear optical elements and photon detectors, so that this scheme is feasible by current technologies. 

\end{abstract}

\maketitle

In the quantum information processing including many quantum protocols and quantum
computation \cite{N-H}, quantum entanglement plays a crucial rule.
Most of the quantum protocols concern with the bipartite
system, mainly because the nature of
multipartite entanglement has not been clarified yet. 
Recently, however, the nature of multipartite entanglement, especially, 
that of tripartite entanglement begins to be clarified.
In \cite{dur00}, D\"ur {\it et  al.} have classified the tripartite 
pure states based on the equivalence under stochastic LOCC (local operations and classical communication).
They showed that there are two different kinds of genuine
tripartite entanglement. One is Greenberger-Horne-Zeilinger (GHZ) states
\cite{ghz89}, which is represented, for example,   as
\begin{eqnarray}
\ket{{\rm{GHZ}}}=\frac{1}{\sqrt{2}}\left(\ket{000}+\ket{111}\right)\,,
\label{eq-ghz}
\end{eqnarray}
where $\{\ket{0}, \ket{1}\}$ is the orthonormal basis for a qubit.
 The other is {\it W} states, which is represented, for example, by
\begin{eqnarray}
\ket{W}=\frac{1}{\sqrt{3}}\left(\ket{001}+\ket{010}+\ket{010}\right)\,.
\label{eq-w}
\end{eqnarray}
These two states cannot be converted to each other by LOCC with nonzero success probability. 
These states show different behavior when one of the qubits is discarded. 
For three qubits in GHZ states, the remaining two qubits are completely unentangled. But, for {\it W} states, the remaining two qubits are still entangled. 
Indeed it was shown that {\it W} states are optimal in the amount of such pairwise entanglement\cite{Koashi00}. 

Many works have been devoted to the study of GHZ states in connection with Bell's theorem \cite{Mer90-1,GHSZ90,Mer90-2,Roy91}, and
violation of Bell inequalities are demonstrated experimentally \cite{zei00}. 
Besides the fundamental studies of GHZ states, several applications of 
these states such as the quantum teleportation \cite{Karlsson01}, 
the quantum secret sharing \cite{hillery99,Cleve99} and the 
quantum key distribution protocol \cite{Kempe99,Dukin01} have been
proposed.

On the other hand, the study of {\it W} states has not been done until recently. For application, the quantum key distribution (QKD) with {\it W} states is proposed \cite{Jaewoo02}, and a {\it W}-class state is used for the optimal universal quantum cloning machine \cite{Buzek96,Buzek97,Gisin97,Buzek97-2,Bruss98,Murao99}.
In fundamental aspects, Cabello \cite{Cabello02} has illustrated some differences between the violation of local realism exhibited by {\it W} states and that by GHZ states. 
The {\it W} states have a clearer prescription for selecting a pair of qubits to be subjected to a Bell's theorem test than the GHZ states have. Thus, not only for the purpose of the realization of some applications, but also for the fundamental interests, it is important to prepare {\it W} states experimentally.

\begin{figure}[tbp]
\begin{center}
 \includegraphics[scale=0.5]{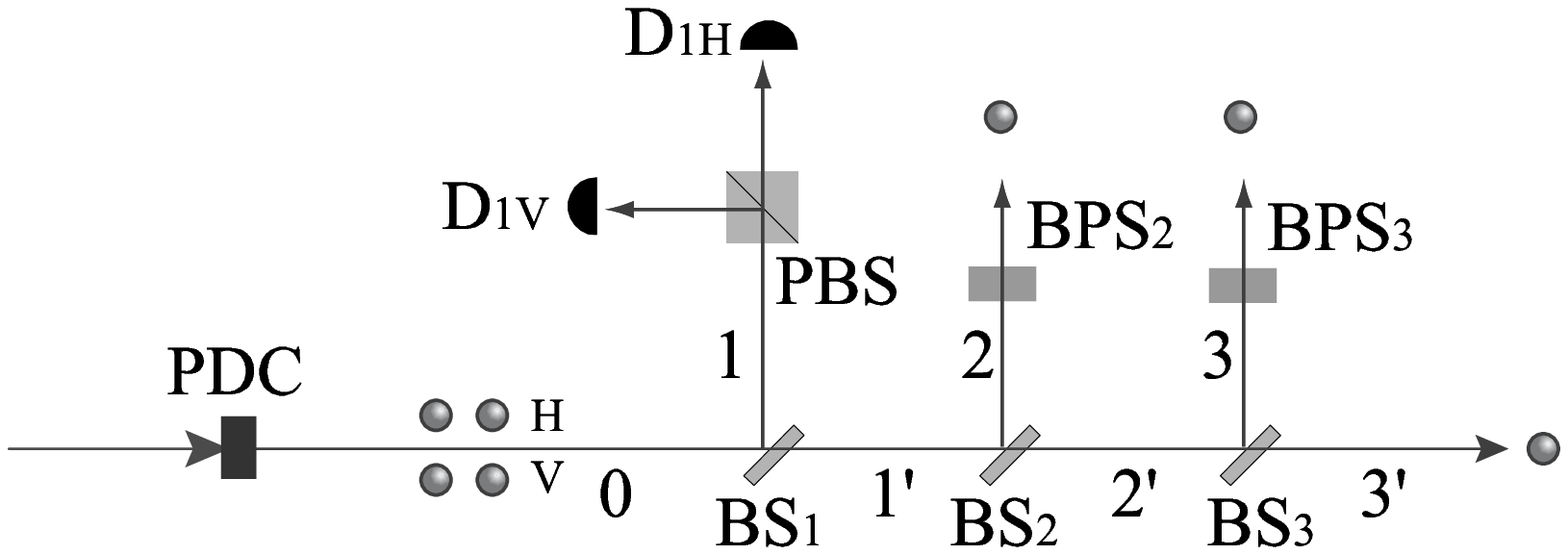}
\end{center}
 \caption
 {\setlength{\baselineskip}{3.5mm} 
The schematic diagram of the setup (scheme I). 
 Polarization beam splitters (PBS) transmit ${\rm H}$ photons and 
 reflect ${\rm V}$ photons. 
}
 \label{fig:setup1}
\end{figure}

So far, several schemes for preparation of {\it W} states have been proposed. Zeilinger, Horne, and Greenberger proposed  a scheme using third order nonlinearity for path entangled photons \cite{Zei97}. Guo and Zhang proposed a scheme for three entangled atoms via cavity quantum electrodynamics \cite{Guo02}. 

In this paper, we propose an experimentally feasible scheme for preparing a polarization entangled {\it W} states. The scheme is composed of parametric down-conversion (PDC), linear optical elements, and photon detectors, so that our scheme is feasible by current technologies. In our proposal, there is no interference
between the photons passing through different paths, which makes it easy for us to align the optical elements and makes our system insensitive to fluctuations of optical path lengths.

In our scheme, we utilize four photons emitted collinearly from type-II PDC, which are in the following state,
\begin{eqnarray}
\ket{2}_{0{\rm H}}\ket{2}_{0{\rm V}},
\label{eq:1}
\end{eqnarray}
where $\ket{n}$ is the normalized $n$-photon number state. 
The subscript numbers label the spatial modes, and ${\rm H}$ and ${\rm V}$ represent horizontal and vertical polarization modes, respectively. As shown in Fig.~\ref{fig:setup1}, these photons are split into four spatial modes (1, 2, 3, and 3$^\prime$) by beam splitters (BS$_k$, $k=1, 2, 3$), whose reflectivity and transmissivity are independent of polarization. The transformation by BS$_k$ is expressed by
\begin{eqnarray}
\ket{1}_{{\rm H}}&\to &r_k\ket{1}_{k{\rm H}}+t_k\ket{1}_{k^\prime {\rm H}}, 
\nonumber \\
\ket{1}_{{\rm V}}&\to &r_ke^{i\phi_{k}}\ket{1}_{k{\rm V}}+t_ke^{i\psi_{k}}\ket{1}_{k^\prime {\rm V}},
\nonumber \\
\ket{2}_{{\rm H}}&\to &r^2_k\ket{2}_{k{\rm H}}+t^2_k\ket{2}_{k^\prime {\rm H}}+2r_kt_k\ket{1}_{k{\rm H}}\ket{1}_{k^\prime {\rm H}}, 
\end{eqnarray}
and
\begin{eqnarray}
\ket{2}_{{\rm V}}&\to &r^2_ke^{2i\phi_{k}}\ket{2}_{k{\rm V}}+t^2_ke^{2i\psi_{k}}\ket{2}_{k^\prime {\rm V}} \nonumber \\
& &+2r_kt_ke^{i(\phi_{k}+\psi_{k})}\ket{1}_{k{\rm V}}\ket{1}_{k^\prime {\rm V}}, 
\end{eqnarray}
where $r_k$ and $t_k$ are the reflection and transmission coefficients of BS$_k$, respectively, which satisfy $|r_{k}|^2+|t_{k}|^2=1$. We assume that $r_k$ and $t_k$ are real, without loss of generality. Here $\phi_{k}$ and $\psi_{k}$ are the phase differences between mode ${\rm H}$ and ${\rm V}$ for reflected and transmitted photons, respectively. For simplicity, we omit the modes in the vacuum, using abbreviations such as $\ket{1}_{k{\rm V}}\ket{0}_{k^\prime{\rm V}} \to \ket{1}_{k{\rm V}}$. After these transformations, the phase offsets for the photons in mode 2 and 3 are compensated by birefringent phase shifters (BPS$_k$, $k=2,3$). The amount of compensation is chosen as 
\begin{eqnarray}
\ket{1}_{{2\rm V}}\to e^{i(-\phi_{2}+\psi_{2}+\psi_{3})}\ket{1}_{{2\rm V}}
\end{eqnarray}
for BPS$_2$, and 
\begin{eqnarray}
\ket{1}_{{3\rm V}}\to e^{i(-\phi_{3}+\psi_{3})}\ket{1}_{{3\rm V}}
\end{eqnarray}
for BPS$_3$.
After compensating these phase differences, we are only interested in the case where there is a single photon in each spatial mode ($1$, $2$, $3$, and $3^\prime$). If such a case is successfully selected, these photons are in the following state, 
\begin{eqnarray}
\frac{1}{\sqrt{2}}(e^{i\phi_{1}}\ket{1}_{1{\rm V}}\ket{W_{{\rm V}}}_{233^\prime}+e^{i(\psi_{1}+\psi_{2}+\psi_{3})}\ket{1}_{1{\rm H}}\ket{W_{{\rm H}}}_{233^\prime})
\label{eq:2}
\end{eqnarray}
where $\ket{W_{{\rm V}}}_{233^\prime}$ and $\ket{W_{{\rm H}}}_{233^\prime}$ are the {\it W} states which can be written as
\begin{eqnarray}
\ket{W_{{\rm V}}}_{233^\prime}&\equiv &\frac{1}{\sqrt{3}}(\ket{1}_{2{\rm H}}\ket{1}_{3H}\ket{1}_{3^\prime {\rm V}}+\ket{1}_{2{\rm H}}\ket{1}_{3{\rm V}}\ket{1}_{3^\prime {\rm H}} \nonumber \\
& &+\ket{1}_{2{\rm V}}\ket{1}_{3{\rm H}}\ket{1}_{3^\prime {\rm H}}) \nonumber 
\end{eqnarray}
and 
\begin{eqnarray}
\ket{W_{{\rm H}}}_{233^\prime}&\equiv &\frac{1}{\sqrt{3}}(\ket{1}_{2{\rm V}}\ket{1}_{3{\rm V}}\ket{1}_{3^\prime {\rm H}}+\ket{1}_{2{\rm V}}\ket{1}_{3{\rm H}}\ket{1}_{3^\prime {\rm V}} \nonumber \\
& &+\ket{1}_{2{\rm H}}\ket{1}_{3{\rm V}}\ket{1}_{3^\prime {\rm V}}). \nonumber
\end{eqnarray}
The probability of obtaining the photons in the state of Eq.~(\ref{eq:2}) is $(2\sqrt{6}r_{1}t^3_{1}r_{2}t^2_{2}r_{3}t_{3})^2$. If we detect a single photon at the photon detector D$_{1{\rm V}}$ and the state is projected to $\ket{1}_{1{\rm V}}\ket{W_{{\rm V}}}_{233^\prime}$, we obtain three photons in the state $\ket{W_{{\rm V}}}_{233^\prime}$. Even if we detect a single photon at the photon detector D$_{1{\rm H}}$ and the state is projected to $\ket{1}_{1{\rm H}}\ket{W_{{\rm H}}}_{233^\prime}$, we can also obtain the state $\ket{W_{{\rm V}}}_{233^\prime}$ after rotating the polarization by $90^{\circ}$ in mode 2, 3, and 3$^\prime$. In this case, the maximum probability of obtaining the photons in the state $\ket{W_{{\rm V}}}_{233^\prime}$ is $3/32$ when we set $r_{1}^2=1/4$, $r_{2}^2=1/3$, and $r_{3}^2=1/2$.
\begin{figure}[tbp]
\begin{center}
 \includegraphics[scale=0.5]{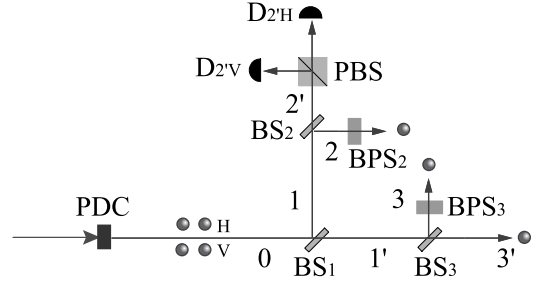}
\end{center}
 \caption
 {\setlength{\baselineskip}{3.5mm}
The schematic diagram of another setup (scheme II). 
 }
 \label{fig:setup2}
\end{figure}
Although it is difficult to select the single photon in each spatial mode without destroying the photons, we can discard the photocounts caused by the non-{\it W} states if we are allowed to perform the postselection where we select the events of the photocounts in mode $2$, $3$, and $3^\prime$. 

In practice, to implement our scheme experimentally, we have to pay attention to the errors and the efficiency of generating the photons in {\it W} states. The
 errors in the selected state are mainly caused by generation of three photon pairs at PDC and the dark counts of photon detectors. In PDC, the photon pair generation rate per pulse $\gamma$ is approximately $10^{-4}$ in typical multi-photon experiments \cite{zei99,zei00,Zeilinger01,Zeilinger02}. The three-pair generation rate $O(\gamma^3)$ is approximately $10^{-4}$ lower than two-pair generation rate $O(\gamma^2)$. The dark counts of current photon detectors is quite low for multi-photon coincidence measurement, so that these errors are negligible. 
(See also \cite{TYamamoto01} about this kind of errors.) To see whether the efficiency of generating three photons in the {\it W} state is acceptable, we compare the yield of the {\it W} state with that of GHZ states in \cite{zei99,zei00,Zeilinger01,Zeilinger02} where type-II PDC is also used for generating three photons. In the GHZ experiment, the probability of obtaining the photons in the GHZ state after generating two photon pairs is $3/8$. Compared with this probability, the yield of {\it W} states in our scheme is smaller by a factor $1/4$ . However, using stimulated PDC \cite{Bouwmeester01}, the four-photon generation rate can be 16 times higher than spontaneous PDC, which suggest that our proposal is experimentally feasible.
 
We can also consider another setup (scheme II) as shown 
in Fig.~\ref{fig:setup2}. 
In this scheme, after compensations similar to scheme I expressed by 
\begin{eqnarray}
\ket{1}_{{2\rm V}}\to e^{i(-\phi_{1}-\phi_{2}+\psi_{1}+\psi_{3})}\ket{1}_{{2\rm V}}
\end{eqnarray}
and 
\begin{eqnarray}
\ket{1}_{{3\rm V}}\to e^{i(-\phi_{3}+\psi_{3})}\ket{1}_{{3\rm V}},
\end{eqnarray}
we obtain the photons in the following state, 
\begin{eqnarray}
\frac{1}{\sqrt{2}}(e^{i\phi_{1}}\ket{1}_{1{\rm V}}\ket{W_{{\rm V}}}_{233^\prime}+e^{i(\psi_{1}-\psi_{2}+\psi_{3})}\ket{1}_{1{\rm H}}\ket{W_{{\rm H}}}_{233^\prime})
\end{eqnarray}
with the probability $(2\sqrt{6}r^2_{1}t^2_{1}r_{2}t_{2}r_{3}t_{3})^2$. The maximum probability of obtaining these photons in the state {\it W} is $3/32$, which is the same as scheme I, when we set $r_{1}^2=r_{2}^2=r_{3}^2=1/2$. This scheme has an advantage that the maximum probability can be obtained by using only symmetric beam splitters.

\begin{figure}[tbp]
\begin{center}
 \includegraphics[scale=0.5]{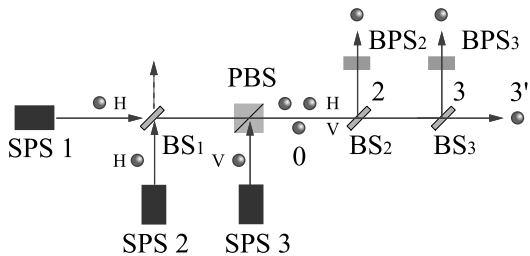}
\end{center}
 \caption
 {\setlength{\baselineskip}{3.5mm}
The schematic diagram of the setup with single photon sources. 
 }
 \label{fig:setup3}
\end{figure}

So far, we have assumed that the reflectivity and transmissivity of BS$_k$ are independent of polarization. If these depend on the polarization, the fidelity of the final state to the desired {\it W} state becomes lower. 
In this case, scheme I and scheme II show slightly different behavior. Here, we represent the polarization-dependent reflection and transmission coefficient of BS$_k$  as $r_{kL}$ and $t_{kL}$, respectively, which satisfy $r_{kL}^2+t_{kL}^2=1$ where $L={\rm H},{\rm V}$. We also introduce the error factor $\delta _{k{\rm L}}$ defined by $\delta _{k{\rm L}}=r_{kL}^2-(r^{opt}_{k})^2$ where $r^{opt}_{k}$ is the optimal reflectivity, 
namely 
$(r^{opt}_{1})^2=1/4$, $(r^{opt}_{2})^2=1/3$, and $(r^{opt}_{3})^2=1/2$ in scheme I and $(r^{opt}_{1})^2=(r^{opt}_{2})^2=(r^{opt}_{3})^2=1/2$ in scheme II. 
When $\delta _{k{\rm L}}$ are small, the fidelity ${\rm F_{I}}$ in scheme I and ${\rm F_{II}}$ in scheme II are given by 
\begin{eqnarray}
{\rm F_{I}}&\approx &1-\frac{1}{24}(27\delta^2_{2}+16\delta^2_{3})+O(\delta _{k}^{3})
\end{eqnarray}
and
\begin{eqnarray}
{\rm F_{II}}&\approx &1-\frac{2}{9}[(2\delta_{1}+\delta_{2})^2 +3\delta^2_{3}]+O(\delta _{k}^{3}).
\end{eqnarray}
where $\delta _{k}=\delta _{k{\rm H}}- \delta _{k{\rm V}}$. 
In scheme I, $\delta _{1}$ merely changes the amplitudes of $\ket{1}_{1{\rm V}}\ket{W_{{\rm V}}}_{233^\prime}$ and $\ket{1}_{1{\rm H}}\ket{W_{{\rm H}}}_{233^\prime}$ in Eq.~(\ref{eq:2}), so that this does not affect the fidelity unlike scheme II.

The use of a single photon source (SPS), which is currently being developed \cite{YYamamoto01,YYamamoto02}, will improve the rate of generating the photons in {\it W} states. An ideal SPS emits a single photon in a single mode at a desired time.  In this case, we can start from only three photons in the state $\ket{2}_{0{\rm H}}\ket{1}_{0{\rm V}}$ (or $\ket{2}_{0{\rm V}}\ket{1}_{0{\rm H}}$ ). To prepare this initial state, three SPSs and a symmetric beam splitter (BS$_1$) are arranged as shown in Fig~\ref{fig:setup3} (SPS1 and SPS2 emit a photon in mode H and SPS3 emits a photon in mode V). The state at the output ports of BS$_1$ is $(\ket{2}\ket{0}+\ket{0}\ket{2})/\sqrt{2}$, so that we obtain three photons in the state $\ket{2}_{0{\rm H}}\ket{1}_{0{\rm V}}$ with probability $1/2$ under the condition that each SPS has emitted a photon.
After we transform these photons by BS$_2$ and BS$_3$, and in the case where there is one photon in each spatial mode, we can obtain the photons in the state $\ket{W_{{\rm V}}}_{233^\prime}$. The probability of obtaining the photons in the state $\ket{W_{{\rm V}}}_{233^\prime}$ after generating a photon from each SPS is  $1/2(\sqrt{6}r_{2}t^2_{2}r_{3}t_{3})^2$ and the maximum of this probability is $3/32$, which is the same as above schemes,  at $r_{2}^2=r_{3}^2=1/2$. 
The generation rate of one photon from SPS is approximately $0.4$ per pulse \cite{YYamamoto02} so that three-photon generation rate is about $0.064$ per pulse, which is significantly larger than $\sim 10^{-8}$ per pulse for PDC \cite{zei99}.  Since SPS and PDC currently achieve almost the same repetition rate, using SPS improves the rate of preparing the state {\it W}. 

In our scheme, one can also prepare non-equally weighted states belonging to {\it W}-class. An example is the state used for the optimal universal quantum cloning machine via teleportation by three distant parties \cite{Bruss98}, 
\begin{eqnarray}
& &\sqrt{\frac{2}{3}}\ket{1}_{2{\rm H}}\ket{1}_{3H}\ket{1}_{3^\prime {\rm V}}
-\frac{1}{\sqrt{6}}\ket{1}_{2{\rm H}}\ket{1}_{3{\rm V}}\ket{1}_{3^\prime {\rm H}} \nonumber \\
& &-\frac{1}{\sqrt{6}}\ket{1}_{2{\rm V}}\ket{1}_{3{\rm H}}\ket{1}_{3^\prime {\rm H}}. 
\end{eqnarray}
To prepare such states, one can generally include additional polarization dependent losses in mode $2$, $3$, and $3^\prime$ and adjust BPS$_k$ properly. 

In summary, we have proposed simple schemes for preparing the the photons in {\it W} states by using parametric down-conversion, linear optical elements, and photon detectors. 
The schemes are easy to implement and feasible by current technologies. Our schemes can be improved by using single photon sources to obtain a higher rate of generating the photons. 


We thank K. Nagata, J. Shimamura, and S. K. \"Ozdemir for helpful discussions.

\end{document}